# Active Learning Through Flexible Collaborative Exams: Improving STEM Assessments

Kristina Callaghan,[1,2] Greg Kestin,[1] Anna Klales,[1] Logan McCarty,[1,3] Louis Deslauriers[1]†

[1]Department of Physics, Harvard University, Cambridge, MA 02138, USA

[2]Department of Physics, Louisiana State University, LA 70803, USA

[3]Department of Chemistry and Chemical Biology, Harvard University, Cambridge, MA 02138, USA

**Abstract**

Two-stage exams, which pair a traditional individual exam with a subsequent collaborative exam, are widely popular with both students and faculty for fostering deep engagement, collaboration, and immediate feedback. Over the last decade, this assessment model has gained substantial traction in STEM college courses; however, holding both stages during class sessions often limits the full potential of two-stage exams, deterring many instructors from adopting them. To accommodate these constraints, instructors must either reduce the exam's content depth and coverage, or hold evening, in-person exam sessions at fixed times, which can often be impractical for students and faculty with external commitments. In this paper, we introduce an alternative asynchronous approach that allows the collaborative portion to be completed unsupervised and outside of regular class hours, within 48 hours of the individual exam. This method not only eases logistical constraints but also allows more time for collaboration, potentially enhancing feedback quality. Our findings show that students equate their engagement, collaboration, and feedback quality during the asynchronous collaborative exam with that of standard in-class exams and show signs of retention of learning. This cost-effective approach requires no extra time or resources and could promote widespread adoption of this effective assessment method, especially in online-only and general STEM courses.

† To whom correspondence should be addressed:

Louis Deslauriers, 17 Oxford Street (Lyman 234), Cambridge, MA 02138, (650) 248-7395, louis@physics.harvard.edu

## I. INTRODUCTION

It is now widely recognized that research-based educational strategies that promote engagement and collaboration among peers substantially enhance both learning outcomes and student attitudes (Crouch & Mazur, 2001; Smith et al., 2009; Deslauriers et al., 2011; Freeman et al., 2014; Deslauriers et al., 2019; Theobald et al., 2020). Two-stage exams, consisting of a conventional individual stage followed by a collaborative group stage, integrate and reinforce collaborative learning within the assessment component of a course. Past studies demonstrate that two-stage exams create intense engagement among students (Rieger & Heiner, 2014; Wieman et al., 2014), promote collaboration (Levy et al., 2023; Leight, 2012; Yu, 2010), provide immediate feedback following an exam (Nicol & Selvaretnam, 2021; Garaschuk & Cytrynbaum, 2019; Gilley & Clarkston, 2014; Rivaz et al., 2015; Vazquez-Garcia, 2018; Vogler & Robinson, 2016; Leight et al., 2012; Rieger & Rieger, 2020), and generally support content retention, with some studies showing retention effects lasting up to seven weeks (Ives, 2014; Cortright et al., 2003; Bloom, 2009; Gilley & Clarkston, 2014). However, these benefits can vary, as not all studies have demonstrated significant long-term retention, and the effects may diminish over time (Kinnear, 2021; Leight et al., 2012; Cao & Porter, 2017).

Although the literature describes the implementation of a two-stage exam strategy as straightforward (Bruno et al., 2017; Gilley & Clarkston, 2014; Rieger & Heiner, 2014; Wieman et al., 2014), the practical challenges of holding two-stage exams within standard class sessions can frequently hinder wider adoption. A typical two-stage exam allocates up to three-quarters of the available exam time to the individual stage and the remaining time to the collaborative stage (Rieger & Heiner, 2014; Rieger & Rieger, 2020). In certain academic settings with longer class times—such as summer courses, once-a-week lecture courses, and extension schools—two-stage exams can be more easily accommodated. However, in most standard courses, which typically feature sessions lasting between 60 to 90 minutes, the time available for each stage is limited. With less time available for the individual stage, instructors face constraints not only in the number and complexity of questions they can pose but also in the breadth of topics covered (Rieger & Heiner, 2014; Rieger & Rieger, 2020; Kinnear, 2021). This limitation can affect both the depth of student understanding and the thoroughness with which the course material is assessed. Research shows that the consensus-finding crucial for learning in the collaborative stage is best stimulated by complex questions that require higher-level cognitive skills (Heller & Hollabaugh, 1992; Stearns, 1996; Rieger & Rieger, 2020; Bloom et al., 1964). Furthermore, adequate time during the individual stage allows students to engage deeply with the material, enabling them to think critically about and commit to their solutions. This preparation is crucial for enhancing their participation in the subsequent collaborative discussions (Wieman et al., 2014).

To mitigate these constraints, instructors sometimes opt to schedule exams outside of regular class hours, such as in extra evening sessions or by extending the start or end times of classes (Rieger & Heiner, 2014; Rieger & Rieger, 2020). While theoretically feasible, especially for the first or last lecture of the day, this strategy often encounters practical barriers such as overlapping class schedules, limited room availability, and conflicts with students' and faculty's external commitments. These challenges are compounded in settings with large class enrollments and are especially pronounced in universities that have a high number of commuter students or those working outside of school, adding a significant administrative burden in the management of make-up exams.

To alleviate these logistical challenges, we introduce an alternative strategy where the individual stage occurs during scheduled lecture sessions, as in a conventional exam, but the collaborative stage is completed asynchronously. This asynchronous approach includes two different formats: an **online** collaborative exam where groups of students meet via conferencing software such as Zoom or an **in-person** collaborative exam where groups of students meet at a location of their choosing, with both formats completed at times chosen by each group. Although the collaborative exams are primarily unsupervised in both formats, instructors remain 'on call' to answer clarifying questions during designated 'office hours,' large blocks of time scheduled based on instructor and teaching assistant availability, thereby effectively balancing student independence with available support. This level of flexibility allows for enhanced student engagement and reduced stress while maintaining some level of academic integrity. The asynchronous model contrasts with the ***standard*** collaborative exam that typically occurs immediately after the individual stage in the classroom and involves all groups participating simultaneously (Rieger & Rieger, 2020; Wieman et al., 2014; Rieger & Heiner, 2014).

In this study, we present comprehensive survey data assessing shifts in students' understanding over time alongside their perceptions of the exam experience following both standard and asynchronous collaborative exams. These evaluations capture students' responses immediately after the exams and again at three-and-a-half and eight-and-a-half weeks later, providing insights into the long-term impact of exam format on learning and engagement. The surveys assessed both satisfaction and perceived learning outcomes and indicate that students perceived the standard and asynchronous collaborative exams as equally effective. Importantly, students' understanding and positive perceptions of the exam process remained consistent across all formats throughout the study period. Our findings suggest that the asynchronous format preserves the educational benefits of two-stage exams—deep engagement, positive student experiences, and enhanced learning retention—without requiring additional in-person class time. These results are particularly relevant for instructors of online courses and those dealing with logistical or administrative constraints when implementing two-stage exams. The flexibility offered by asynchronous exams can promote broader adoption of this highly effective assessment method, as evidenced by its successful implementation in several courses across two institutions.

## II. METHODS

**Course context**

The investigation was conducted in a quantum mechanics course at Harvard University, covering standard topics typically found in any junior-level quantum mechanics curriculum. As a core part of the physics department's offerings, this course attracts a broad spectrum of students and typically enrolls around 60 students each semester. Lectures are held twice per week, each lasting 75 minutes, and are delivered in an interactive lecture style. This style incorporates small-group work interspersed with direct feedback from an instructor experienced in active learning methodologies (Deslauriers et al., 2011; Deslauriers & Wieman, 2011; Jones et al., 2015; McCarty & Deslauriers, 2020; Dunleavy et al., 2022). Discussion sections, held weekly, can be viewed as an extension of the lecture, providing more practical applications in a smaller, interactive setting of 15–20 students. Homework assignments are designed to encourage deliberate practice, focusing on developing specific subskills with detailed and immediate feedback (Miller et al., 2021). The semester spans 15 weeks and includes two scheduled midterm exams in the 5th and 11th week.

Approximately half of the students were physics majors, while the remainder pursued degrees in engineering, astrophysics, applied mathematics, and life sciences. Despite being a junior-level course, two-thirds of the class were first and second-year students, with the remainder being juniors and seniors, creating a diverse mix of students. In this particular iteration of the course, 55 were enrolled.

**Experimental design**

The experimental design of this study spanned two midterm exams, each consisting of three complex long-answer problems requiring multi-step problem-solving and one problem comprised of several multiple-choice questions, each of which addressed a conceptual idea. The individual exams were administered in-class over 80 minutes. Midterm 1 served as the control condition with an in-person collaborative exam to ensure all students had experienced the standard two-stage exam format. Due to the course's midday timing, the collaborative exam session was scheduled in three separate time slots later the same day to accommodate students' schedules. Even with three separate time choices, there were still six students who continued to have scheduling conflicts. Students discussed the exam content in self-selected groups of three or four, addressing the same problems from the individual exam along with an additional bonus problem designed to deepen engagement and application of the topics. A student's overall score on the midterm was calculated as a weighted average, with 80% of the score based on the individual exam and the remaining 20% based on the group score; group scores could only help a student's grade (i.e., if the group score was less than the individual score, the individual score made up 100% of the overall score) and all members of a group received the same group score. This 80% individual and 20% group partition is a common weighting scheme and reflects standard practices in the field (Wieman & Rieger, 2014; Rieger & Rieger, 2020). Allotting 80% of the overall score based on a student's individual performance also increases individual accountability and lessens the likelihood of any issues of free-loading, a possible but uncommon occurrence during collaborative exams (Levy et al., 2023; Rieger & Heiner, 2014; Zipp, 2007).

In contrast, Midterm 2 introduced the experimental condition by varying the collaborative stage format; the self-selected groups from Midterm 1 were randomly assigned to either collaborate via Zoom ('asynchronous online' exam) or meet physically ('asynchronous in-person' exam). The presence of two asynchronous formats was designed to investigate whether students could be afforded maximum flexibility for the collaborative stage—choosing when and where to meet for in-person groups and when to connect online. This level of flexibility, particularly the option of online collaboration, is important not only for instructors at non-residential colleges whose students may find it challenging to be physically present on campus but also for instructors of entirely online courses, where coordinating in-person meetings is inherently impractical. Despite the different asynchronous formats, both groups experienced the collaborative stage under identical constraints: unsupervised, untimed, closed-book, and submitted electronically within 48 hours. Overall scores on Midterm 2 were calculated as in Midterm 1.

Immediately after each exam component – individual and collaborative, – 'understanding self-reflections' were administered to gauge how students' perceived grasp of the material changed over time. This consisted of asking students to answer the question "Now that I have worked on the exam, I am now confident that I understand the solution to _____ problems out of 4." Answer choices for the understanding self-reflection ranged from "All 4 problems" to "None." Additionally, surveys assessing students' perceptions, attitudes, and preferences were conducted after each collaborative exam to evaluate the impact of the different exam formats. Within these surveys, students were also invited to include open-ended comments about their experience. Every student enrolled in the course (N = 55) participated in the surveys. Given that this study involved standard educational practices within normal classroom settings, it was exempt from Institutional Review Board (IRB) oversight. Figure 1 summarizes these assessments within the midterm exam structure.

**Post-Understanding Self-Reflection**

In the semester's final lecture, students engaged in a structured review of both midterm exams. The students spent approximately 10 minutes working individually on each problem — four from each midterm —-and were instructed to focus primarily on a problem-solving approach for each question instead of numerical calculations, thereby ensuring they prioritized conceptual understanding over tedious computations. Full participation credit was awarded based on effort, thereby encouraging meaningful contributions regardless of whether the student completed all calculations. This structured review was specifically designed to reassess students' mastery of complex problem-solving tasks seen on previous exams and not just their subjective perception of learning.  This approach not only provided a meaningful review for students but also allowed for the administration of a post-understanding self-reflection in a contextually relevant manner, immediately after they engaged with the exam content.

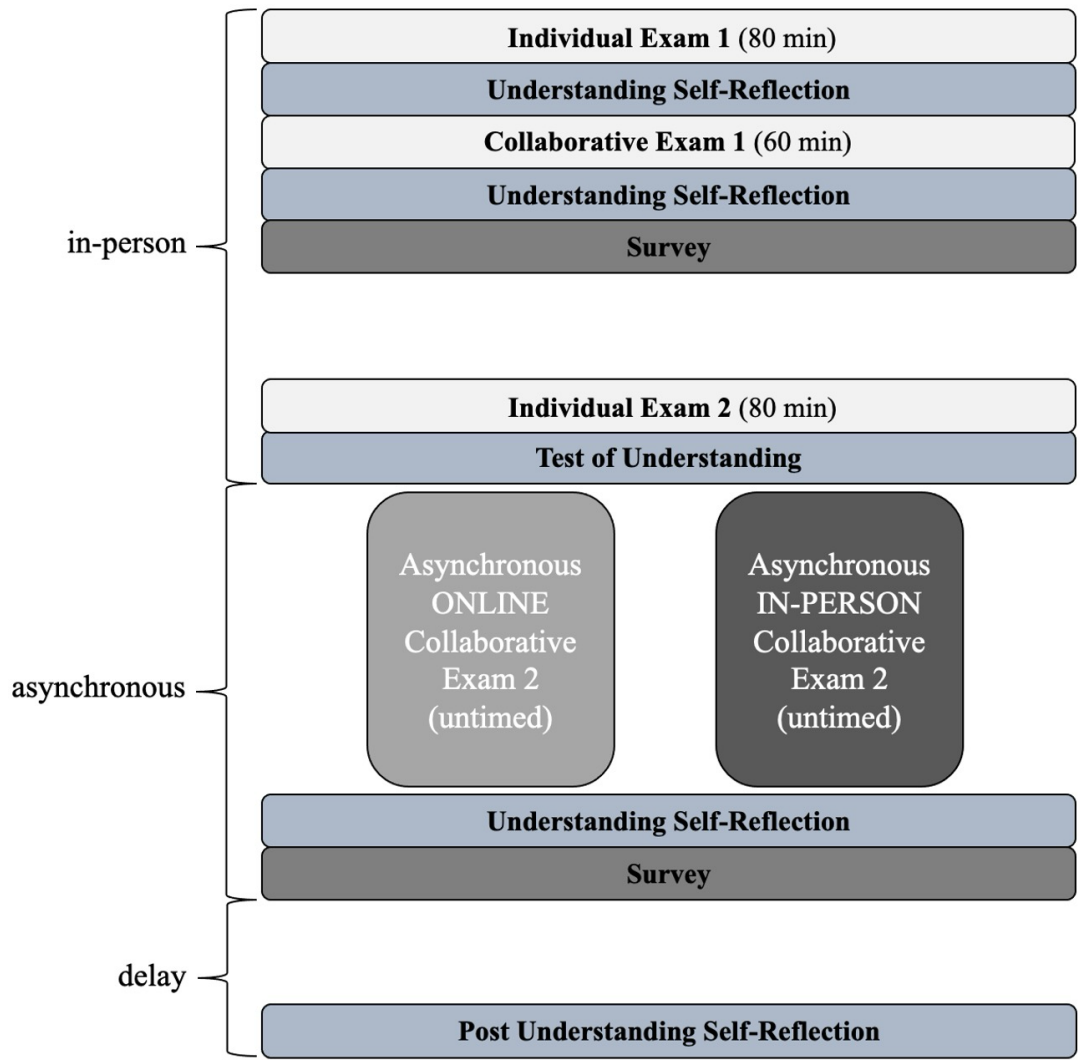

**Figure 1.** Overview of the midterm exams, surveys, and understanding self-reflections, administered after each exam stage to assess students' grasp of the material and their attitudes/satisfaction with the exam format. The post-understanding self-reflection included a structured review session for each midterm question and took place 8.5 weeks after Midterm 1 and 3.5 weeks after Midterm 2.

## III. RESULTS

The Results section presents students' perceptions of the collaborative exam across standard and asynchronous formats. For all Likert-scale questions reported in Figures 2 – 4, the distributions of responses were approximately normal, with no extreme skewness or outliers. Group means are therefore used to summarize trends across conditions, with error bars indicating one standard error of the mean. All statistical tests used a significance level of $\alpha$ = 0.05. Given the focus of this study on pre-planned comparisons, corrections for multiple comparisons were not applied. These analyses aim to demonstrate that both asynchronous approaches provide experiences consistent with the more rigid, in-person group exams, supporting their adoption as a convenient alternative rather than focusing on potential minor differences between the conditions.

Figure 2 illustrates students' perceptions of the collaborative exam, measuring their enjoyment, perceived learning, mental engagement, and the usefulness of the feedback. The responses, captured on a 5-point Likert scale, reveal no significant differences between the standard and asynchronous formats in terms of student satisfaction and overall response, confirming the consistency of these perceptions.

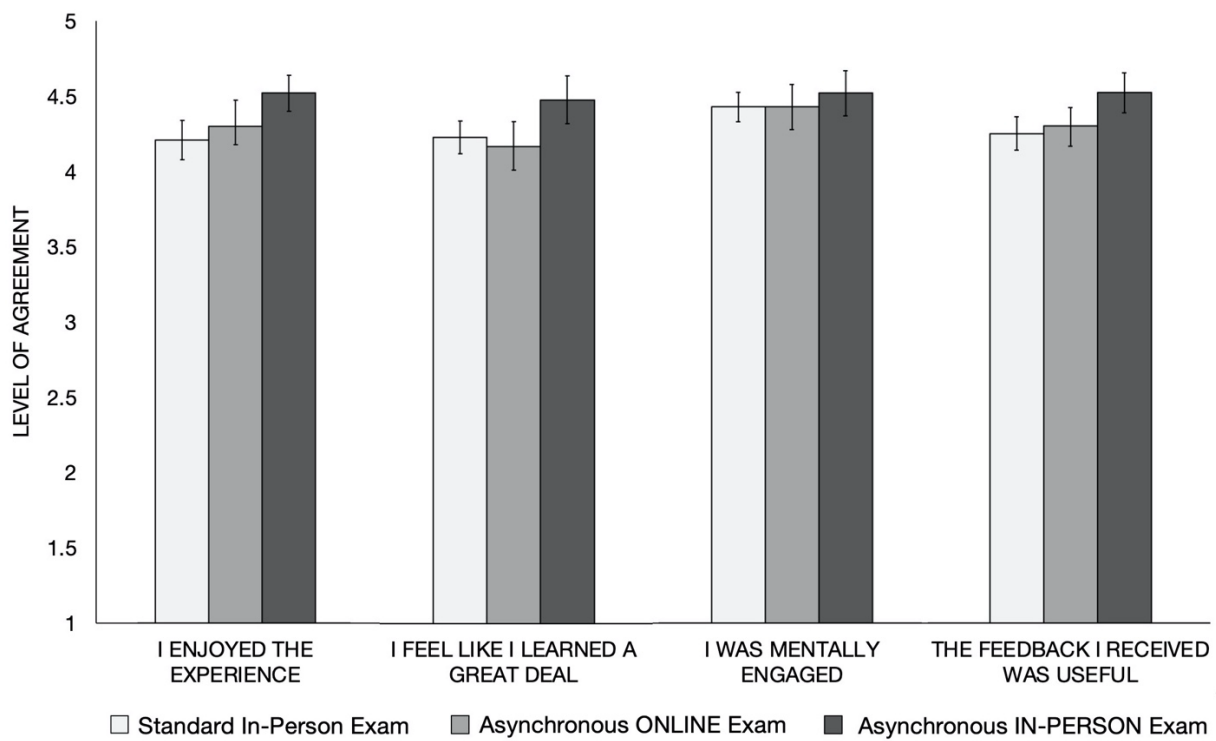

**Figure 2.** Survey results comparing students' perceptions about the collaborative stage as a function of exam format. All students completed a standard, in-person collaborative exam during Midterm 1. During the collaborative stage of Midterm 2, students completed the exam asynchronously either by meeting using Zoom or in-person. No statistical differences were seen between student responses across the collaborative exam implementations. Error bars indicate

one standard error from the mean. Distributions were approximately normal, with no significant skewness. The full text for each question, from left to right, was: "I enjoyed the group exam experience;" "I feel like I learned a great deal from the group exam;" "I was mentally engaged during the group exam;" and "The feedback I received during the group exam was useful."

After the Midterm 2 collaborative exam, students were invited to directly compare the asynchronous exam experience with the standard in-person collaborative exam of Midterm 1. This prior exposure to the standard, in-person collaborative exam provided a meaningful reference point, allowing students to evaluate the asynchronous format against a familiar setup they had previously experienced.

We were particularly interested in whether the flexibility of scheduling and the removal of time constraints in the asynchronous format would enhance collaboration quality and further reduce exam-related anxiety, as anxiety reduction is known to be a benefit of standard two-stage exams (Rempel et al., 2021; Fournier et al., 2017; Pandey & Kapitanoff, 2011; Kapitanoff, 2009). Additionally, we aimed to assess the impact of reduced instructor oversight on the exam process. Figure 3 presents student responses to these aspects. Students generally reported less anxiety with the asynchronous exam compared to the standard setup. An independent-samples t-test comparing the anxiety levels of students in the asynchronous online and in-person formats revealed a statistically significant difference, $t(55) = 2.08$, $p < 0.04$. Students in the in-person format (M = 4.6, SE = 0.16) reported slightly lower anxiety compared to those in the online format (M = 4.1, SE = 0.18). Although slightly lower, the online score still indicates agreement, confirming that students responded favorably to the untimed nature and the flexibility to schedule the exam at a convenient time. Both cohorts also demonstrated agreement that removing the time constraint on the asynchronous exam led to increased collaborations and discussion (online: 4.1 $\pm$ 0.2 versus in-person: 4.48 $\pm$ 0.16, $p < 0.10$). Regarding the role of instructor availability, responses were moderately neutral, suggesting that the unsupervised nature of the asynchronous exam did not critically undermine its effectiveness (online: 3.8 $\pm$ 0.2 versus in-person: 3.6 $\pm$ 0.2, $p < 0.4$).

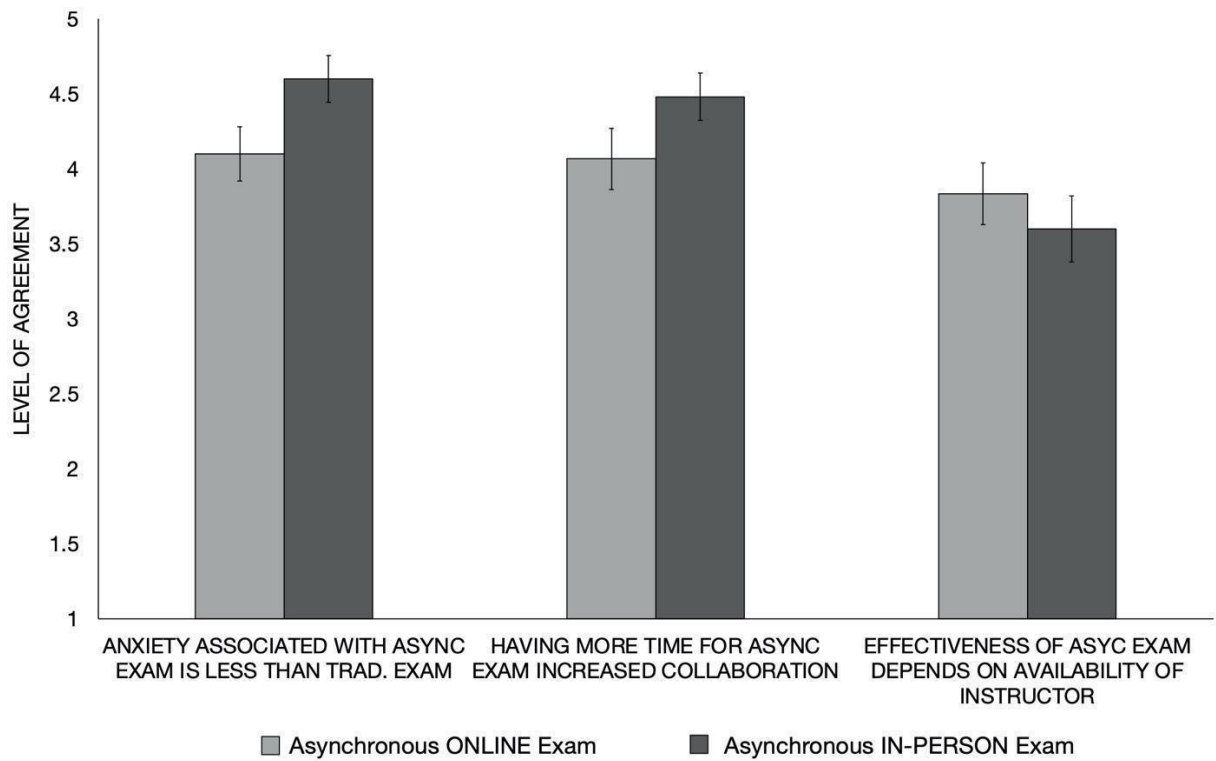

**Figure 3.** Survey results comparing anxiety levels, collaboration quality, and the need for an instructor in asynchronous versus standard collaborative exams. Scores reflect agreement on reduced anxiety and enhanced collaboration without time constraints, for both online and in-person formats. Moderately neutral responses regarding the importance of instructor availability are also noted. The full text for each question, from left to right, was: "The overall anxiety associated with a remote (online) group exam is less than with the traditional (timed) in-person group exam;" "Compared with "traditional" group exams, having more time to complete the remote (online) group exam led to more discussions and collaborations;" and "The effectiveness of the remote (online) group exam format critically depends on being able to text the instructor for help." Error bars represent one standard error from the mean. Distributions were approximately normal, with no significant skewness.

We additionally explored how the online exam format influenced collaborative sharing of work. Figure 4 presents data on the ease with which students who completed the online collaborative exam engaged in collaborative discussions for multiple choice (MC) and long, written response (Long Ans.) questions online. The results indicate that students found it quite feasible to share and collaborate on multiple choice (4.47 ± 0.15) and long-answer questions (4.13 ± 0.17), suggesting the type of question had minimal impact. In fact, when directly asked if it had been easier to share work on multiple choice questions compared to long answer questions, students overall responded neutrally (3.60 ± 0.18). This ease of collaboration is reflective of the students' proficiency with digital tools (Stoian et al., 2022; Hollister et al., 2022), a likely consequence of the widespread adoption of platforms like Zoom during the pandemic.

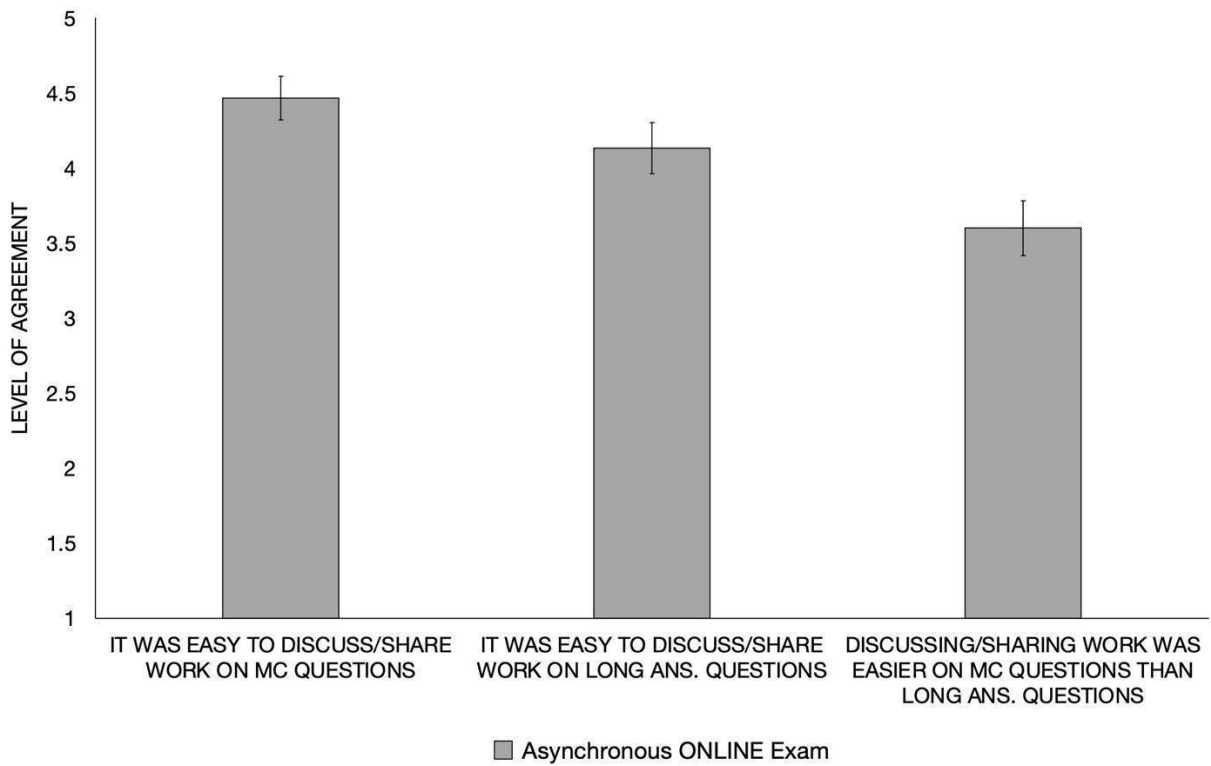

**Figure 4.** Survey results on student collaboration during the asynchronous online exam, specifically regarding the ease of discussing and sharing multiple choice (MC) and long, written response (Long Ans.) questions. The full text for each question, from left to right, was: It was easy to discuss/share work on Multiple Choice questions;" "It was easy to discuss/share work on Long Answer problems;" and "Discussing/sharing work was easier on Multiple Choice questions

than on Long Answer problems." Error bars reflect one standard error of the mean. Distributions were approximately normal, with no significant skewness.

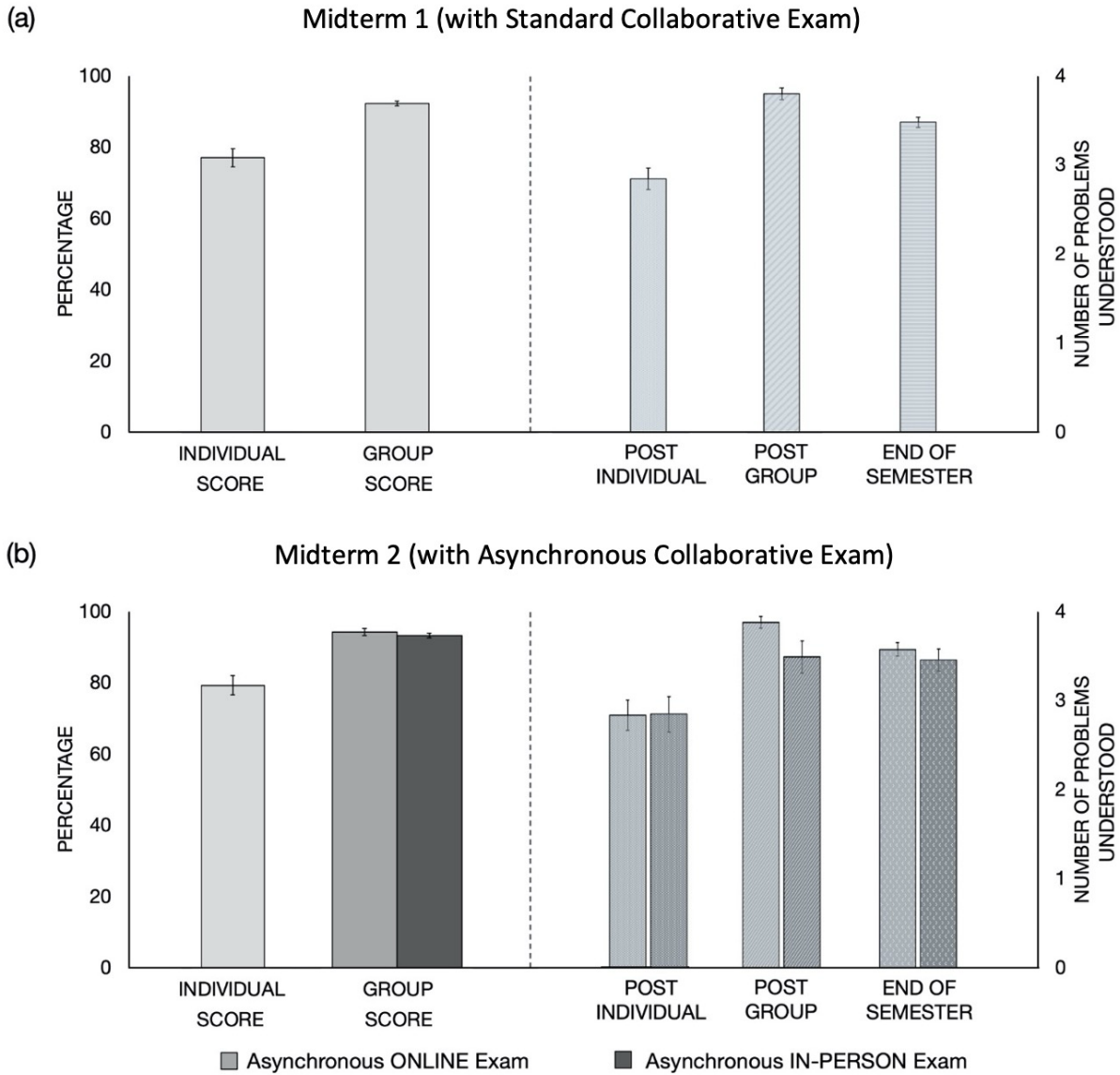

**Figure 5.** Part (a) and part (b) display the average scores for the individual and collaborative exams of Midterm 1 and Midterm 2, respectively, alongside understanding self-reflections conducted on three separate occasions: immediately post-individual exams, post-group exams, and after the end-of-semester structured review activity. A dotted line distinguishes actual exam scores on the left from the understanding self-reflection scores on the right. As each midterm consisted of four questions, responses to the understanding self-reflections are reported out of four. Midterm 1 was conducted using a standard, in-person collaborative exam, while Midterm 2 utilized asynchronous formats, allowing for a comparative analysis of these educational methods.

Lastly, Figure 5 provides a comprehensive overview of the outcomes from both midterms and the understanding self-reflections. Notably, students scored higher on the collaborative exam compared to the individual exam, regardless of the format for the collaborative stage. As described in the Methods section, the understanding self-reflection consisted of asking students the question, "Now that I have worked on the exam, I am now confident that I understand the solution to _____ problems out of 4" after each exam stage and after the end-of-semester structured review. Answer choices for the understanding self-reflection ranged from "All 4 problems" to "None." Overall, while no significant differences were observed for most comparisons between exam formats, the online cohort from Midterm 2 perceived a higher level of understanding following the collaborative exam (3.87 ± 0.07 out of 4 problems) in contrast to the in-person group (3.48 ± 0.18 out of 4 problems), with a p-value less than 0.05. Overall, students' self-reported understanding of the exam questions was consistent with their actual exam scores following the individual and group exams, for both midterms and all types of group exams.

**IV. DISCUSSION**

Our study shows that it is feasible to transition the collaborative component of a two-stage exam to an asynchronous format—conducted outside of scheduled class times, either online or in-person—without sacrificing student satisfaction or learning outcomes. This alternative approach alleviates the scheduling and logistical challenges that have traditionally hindered the widespread adoption of two-stage exams, making this cost-effective and flexible assessment strategy more accessible to a broader range of instructors, particularly those of online-only and various STEM courses. Moreover, the overwhelming majority of students expressed a strong preference for scheduling their collaborative exams at times that suited them best. Survey data revealed that 84% of students favored the option to schedule the exam asynchronously at their convenience over fixed, instructor-scheduled times in the evening. This preference underscores the fact that many students have a variety of activities and responsibilities that make it inconvenient for them to attend a collaborative exam outside of regular class hours. Furthermore, survey results confirm that asynchronous collaborative exams effectively sustain student engagement and enjoyment while providing meaningful feedback, as shown in Figure 2.

During the asynchronous collaborative exams, casual instructor observations of asynchronous in-person groups gathered near instructor offices and brief, informal check-ins during online sessions revealed the same level of intense student engagement typically observed in standard in-class collaborative exam settings. In fact, one instructor, having observed the intense engagement among students, remarked that students were so engrossed in discussion that they seemed unaware of his presence. These observations indicate that the asynchronous formats effectively replicate the essential interactive component of collaborative exams.

Having observed these trends, we sought to understand the underlying reasons why these asynchronous formats could successfully replicate the benefits typically seen in standard, in-class two-stage exams. According to the literature (Rieger & Heiner, 2014; Wieman et al., 2014), two-stage exams excel for two primary reasons: (i) exams are high-stakes assessments and (ii) students are well-prepared to engage in group work after the time spent studying for the exam and after thinking carefully and committing to answers during the individual stage. Our asynchronous format maintained these conditions without any alteration to the exam's core structure and only removed the constraints of specified time and physical location. While one might think students could use the easing of instructor oversight to search for solutions using outside resources during an asynchronous collaborative exam, the literature suggests that even during open-book two-stage exams, students discuss the exam rather than search for answers (Rieger & Heiner, 2014; Hsu, 2021). Student comments reflect that this also occurred during the asynchronous exam and stated that groups spent the extra allowed time ensuring all group members agreed upon and understood each problem before moving on.

Furthermore, removing the time constraint inherent to a standard collaborative exam appears to have had a positive effect. Specifically, students in both asynchronous cohorts reported an increase in the quality of their collaborations and an overall decrease in exam anxiety. When asked how long it took for them to complete the collaborative exam, students in both the online and in-person asynchronous cohorts reported spending an average of approximately 1.6 hours on the exam, which is unsustainable in most standard two-stage exam environments. Not having a member of the teaching staff immediately available did not seem to be a significant problem for students. During the asynchronous exam period, 2 out of 9 groups in the online cohort and 3 out of 10 in the in-person cohort reached out with questions. The nature of these questions mirrored typical standard, in-person collaborative exams and were primarily short, clarifying questions. The instructors' interactions were brief, with an instructor on-call spending approximately 20 minutes in total over 48 hours responding to questions for a class of 55 students and 19 groups. Lastly, the students who completed the asynchronous online exam were able to collaborate and share work effectively, suggesting they possessed a digital fluency and resiliency that makes an online asynchronous exam possible without losing the previously reported benefits of two-stage exams. This might not have been possible before the widespread adoption of remote learning tools during the COVID-19 pandemic.

**Academic integrity**

The shift to asynchronous exams, while enhancing flexibility and student engagement, also introduces potential challenges, particularly regarding academic integrity. The reduced direct supervision in asynchronous settings can raise concerns about potential academic dishonesty. However, although we did not actively investigate the occurrence of academic dishonesty, we also did not observe any evidence of increased cheating, which supports the notion that students are more focused on genuine collaboration during collaborative exams rather than searching for easy answers (Rieger & Heiner, 2014; Hsu, 2021). To further mitigate these risks, it is crucial for instructors to remain proactive. This includes being available for brief questions and designing exams that emphasize collective problem-solving over mere answer sharing. However, it is important for instructors to also remember that one of the main reasons for holding a collaborative exam is to provide students with immediate feedback and an opportunity for collaborative learning; the purpose of the collaborative exam is not solely to assess students.

**Retention of learning**

Our results suggest that an asynchronous collaborative exam can offer feedback as useful as that provided in standard, in-person settings. Student responses to the understanding self-reflection question 'Now that I have worked on the exam, I am now confident that I understand the solution to _____ problems out of 4' are particularly revealing in this regard and were self-calibrated by asking this same question on three separate occasions. Comparing student responses immediately following the individual stage to the actual exam averages shows that while students slightly underestimate their scores, they demonstrate their understanding on a binary scale—either confident or not, for each problem—which reduces ambiguity and provides a more definitive measure than typical subjective perceptions.

The collaborative exams functioned as expected, resulting in both higher actual scores and higher reported understanding than after the individual exam alone, across all exam modalities: standard, asynchronous online, and asynchronous in-person. Interestingly, the online asynchronous exam cohort reported a higher level of understanding than those in the in-person asynchronous cohort. While the reasons for this difference remain uncertain, possible factors could include the unique group dynamics fostered by online interaction, such as students feeling less inhibited and more willing to express their opinions online (Hollister et al., 2022). Furthermore, in the post-understanding self-reflection, students reported their level of understanding as lower than immediately after the collaborative exam but significantly higher than after the individual exam. This consistent understanding self-report across both asynchronous modalities and midterms underscores the effectiveness of the asynchronous format in potentially enhancing student learning outcomes.

**Diverse settings**

The convenience and flexibility of asynchronous two-stage exams have led to their increased adoption in various courses. In two introductory physics courses at the University of California Merced, one of which was entirely online, we informally surveyed students using the satisfaction survey outlined in Figure 2 of this study. This institution, designated as R2 by the Carnegie Classification system, serves a diverse student body with 59% Pell-grant eligible and 65% first-generation college students (University of California Merced, n.d.). The uniformly positive feedback received suggests that the successes of the asynchronous exam are valid across different student populations and are not contingent on prior experience with in-person standard group exams. Notably, the high satisfaction observed in the fully online course demonstrates that online asynchronous group exams are robust, effectively engaging students even without prior in-person collaborative exam experience. In these UC Merced courses, as well as additional courses at Harvard, all midterm exams used the asynchronous two-stage exam strategy and students never experienced a standard, in-person collaborative exam. Student feedback remained consistently positive, though, mirroring the satisfaction levels observed in this study. These findings suggest that the asynchronous format is robust and effective across various educational environments and student demographics, thereby enhancing its potential for wider adoption.

**Recommendations for instructors**

Drawing from the extensive experience of implementing asynchronous two-stage exams across various courses, we provide key suggestions to ensure their successful implementation in Table 1. Our recommendations are aimed at effectively coordinating both stages of the exam and fostering a productive collaborative environment. These include adopting strategies that provide the right mix of flexibility, structure, and support during the exam period. A 'Sample Message to Students' provided in the Supplemental Materials outlines the rationale and procedure for asynchronous two-stage exams and establishes clear expectations to encourage student participation and accountability.

| Key factors for successful asynchronous implementation | Description |
|---|---|
| Student Onboarding | Explain to students the benefits of two-stage exams and the rationale for holding the collaborative stage outside of class. More details can be found in the Supplemental Materials. |
| Exam Question Type and Difficulty | Include complex, conceptual questions and avoid questions that require only factual recall. The difficulty of the questions should be such that a high score is obtainable with a group of 3 or 4 students. |
| Grading Scheme | Allocate 15% - 20% of the overall exam score to the collaborative stage—this incentivizes and rewards students for productive collaboration and effort during the collaborative stage. |
| Group Composition | Restrict group sizes to 4 to ensure participation from all students; accommodate pre-existing groups and assign ungrouped students intentionally to ensure a mix of preparedness levels for balanced collaboration. |
| Group Communication | Form groups at least one week prior to the exam and encourage students to obtain contact information from all group members. Send a coordination email to each group to ensure students know who is in their group. |
| Straightforward Communication and Policies | Send announcements through the course learning management system announcing exam policies, deadlines, and submission reminders. Have a policy in place to deal with students who are unreachable during the collaborative exam. Hold "exam office hours" |

> so students can communicate with the instructor during the exam. Advise students on roughly how long they should expect to spend on the exam and include an upper time limit (i.e., no more than 3 hours).

---

**Table 1.** Key recommendations for successfully implementing an asynchronous collaborative stage. Additional details, including sample text for instructor communication with students, are provided in the Supplemental Materials.

## V. CONCLUSION

In conclusion, our findings suggest that an asynchronous collaborative portion of a two-stage exam can achieve effectiveness comparable to that of standard in-person formats. The type of asynchronous implementation—whether in-person or online—appears to be a secondary consideration, as students overwhelmingly favor the flexibility of asynchronous exams. This flexibility in asynchronous format not only makes two-stage exams more accessible across a wider array of STEM courses but also easier to implement, leading to the rapidly increasing adoption of asynchronous two-stage exams.

Moreover, the asynchronous model also allows instructors to fully utilize class time for individual assessments and gives students the freedom to engage more deeply and thoughtfully in group discussions, which can alleviate anxiety and enhance learning. However, this model requires careful management of academic integrity due to the reduced direct supervision. We recommend that instructors remain accessible for student questions and design the exams to promote genuine collaboration, focusing on the formative value of group interactions over strict testing.

The consistency of positive outcomes across various educational settings suggests that these benefits are likely to extend broadly. This approach allows instructors to emphasize the formative nature of the group exam and treat it as an opportunity for deep learning, rather than merely as an assessment tool.

**Acknowledgments**

The authors acknowledge valuable discussions with Carlos Argüelles-Delgado and Melissa Franklin.

**Disclosure statement**